\documentclass[11pt]{article}
\usepackage{fullpage}
\usepackage{geometry}                % See geometry.pdf to learn the layout options. There are lots.
\usepackage{graphicx}
\usepackage{amssymb}
\usepackage{amsmath}
\usepackage{amsthm}
\usepackage{MnSymbol}
\usepackage{hyperref}
\def\e{{\rm e}}
\def\ii{{\rm i}}
\def\be{\begin{equation}}
\def\ee{\end{equation}}

\def\x{x_{\rm c}}

\newtheorem{definition}{Definition}
\newtheorem{thm}{Theorem}
\newtheorem{prop}{Proposition}
\newtheorem{lemma}{Lemma}

\usepackage[font=small, labelfont=bf, width=0.8\textwidth]{caption}

\title{Off-critical parafermions and the winding angle distribution of the O($n$) model}
\author{Andrew Elvey Price, Jan de Gier, Anthony J Guttmann and Alexander Lee\\ \ \\ {\small Department of Mathematics and Statistics}\\{\small The University of Melbourne, VIC 3010, Australia}}
\date{\today}                                           % Activate to display a given date or no date

\begin{document}
\maketitle

\abstract{
Using an off-critical deformation of the identity of Duminil-Copin and Smirnov, we prove a relationship between half-plane surface critical exponents $\gamma_1$ and $\gamma_{11}$ as well as wedge critical exponents $\gamma_2(\alpha)$ and $\gamma_{21}(\alpha)$ and the exponent characterising the winding angle distribution of the O($n$) model in the half-plane, or more generally in a wedge of wedge-angle $\alpha.$ We assume only the existence of these exponents and, for some values of $n,$ the conjectured value of the critical point. If we assume their values as predicted by conformal field theory, one gets complete agreement with the conjectured winding angle distribution, as obtained by CFT and Coulomb gas arguments. We also prove the exponent inequality $\gamma_1-\gamma_{11} \ge 1,$ and its extension $\gamma_2(\alpha)-\gamma_{21}(\alpha) \ge 1$ for the edge exponents. We provide conjectured values for all exponents for $n \in [-2,2).$
\let\thefootnote\relax\footnotetext{Email: \texttt{andrewelveyprice@gmail.com, jdgier@unimelb.edu.au,\\  t.guttmann@ms.unimelb.edu.au, a.lee19@pgrad.unimelb.edu.au}}
}

\section{Introduction}
The $n$-vector model, introduced by Stanley in 1968 \cite{S68} is described by the Hamiltonian $${\mathcal H}(d,n) = -J\sum_{\langle i,j \rangle} {\bf s_i }\cdot{\bf  s_j} ,$$
where $d$ denotes the dimensionality of the lattice, and ${\bf s_i}$
is an $n$-dimensional vector of magnitude $\sqrt{n}$. When $n=1$ this Hamiltonian
describes the Ising model, when $n=2$ it describes the classical
XY model, and in the limit $n \to 0,$ one recovers the self-avoiding walk (SAW) model, as first pointed out by de Gennes \cite{dG72}. The $n$-vector model has been shown to be equivalent to a loop model with a weight $n$ assigned to each closed loop \cite{DMNS81} and weight $x$ to each edge of the loop. The two-dimensional O($n$) model on a honeycomb lattice, which is the focus of this paper, is a particular case of this equivalence. The partition function of the loop model can be written as
\be
Z(x)=\sum_{G}x^{l(G)}n^{c(G)},
\ee
where $G$ is a configuration of loops, $l(G)$ is the number of loop segments and $c(G)$ is the number of closed loops. The parameter $x$ is defined by the high-temperature expansion of the $O(n)$ model partition function and is related to the coupling $J$, the temperature $T$ and Boltzmann's constant $k$ by 
\be
\e^{ \frac{J}{kT} {\bf s_i }\cdot{\bf  s_j} } \approx 1+ x{\bf s_i }\cdot{\bf  s_j}.
\ee

In 1982 Nienhuis \cite{N82} showed that, for $n \in [-2,2],$ the model on the honeycomb lattice could be mapped onto a solid-on-solid model, from which he was able to derive the critical points and critical exponents, subject to some plausible assumptions. These results agreed with the known exponents and critical point for the Ising model, and predicted exact values for those models corresponding to other values of the spin dimensionality $n.$ In particular, for $n=0$ the critical point for the honeycomb lattice SAW model was predicted to be $x_{\rm c}=1/\sqrt{2+\sqrt{2}},$ a result finally proved 28 years later by Duminil-Copin and Smirnov \cite{DC-S10}. 

The proof of Duminil-Copin and Smirnov involves the use of a non-local \emph{parafermionic observable} $F(z)$ where $z$ is the (complex) coordinate of the plane. This function can be thought of as a complex function with the ``parafermionic'' property $F(\e^{2\pi\ii} z) = \e^{-2\pi\ii \sigma} F(z)$ where the real-valued parameter $\sigma$ is called the $\emph{spin}$. For special values of $\sigma$, this observable satisfies a discrete analogue of (one half of) the Cauchy-Riemann equations. This \emph{discrete holomorphic} or \emph{preholomorphic} property allowed Smirnov and Duminil-Copin to derive an important identity for self-avoiding walks on the honeycomb lattice and, consequently, the Nienhuis prediction for $x_{\rm c}$.

Smirnov~\cite{Smirnov10} has also derived such an identity for the general honeycomb $O(n)$ model with $n \in [-2,2]$. This identity provides an alternative way of predicting the value of the critical point $x_{\rm c}(n)=1/\sqrt{2+\sqrt{2-n}}$ as conjectured by Nienhuis for values of $n$ other than $n=0$.

This paper contains two new results. We first present an off-critical deformation of the discrete Cauchy-Riemann equations, by relaxing the preholomorphicity condition, which allows us to consider critical exponents near criticality. Indeed, this deformation gives rise to an identity between bulk and boundary generating functions, and we utilize this identity in Section~\ref{ssec:wedge} to determine, based on some assumptions, the asymptotic form of the winding angle distribution function for SAWs on the half-plane and in a wedge in terms of boundary critical exponents. It is important to note that up to this point the only assumptions we make are the existence of the critical exponents and the value of the critical point. We will not rely on Coulomb gas techniques or conformal invariance. We find perfect agreement with the conjectured winding angle distribution function on the cylinder predicted by Duplantier and Saleur \cite{DS88} in terms of bulk critical exponents. Finally we conjecture the values of the wedge critical exponents as a function of the wedge angle for $n\in [-2, 2)$.

%%%%%%%%%%%%%%%%%%%%%%%%%%%%%%%%%%%%%%%%%%%%%%%%%%%%%%%%%%%%%%%%
\section{Off-critical identity for the honeycomb O($n$) model}
%%%%%%%%%%%%%%%%%%%%%%%%%%%%%%%%%%%%%%%%%%%%%%%%%%%%%%%%%%%%%%%%
\subsection{Smirnov's observable on the honeycomb lattice}
We briefly review an important result of Smirnov for self-avoiding walks on the honeycomb lattice. 

Firstly, a $\emph{mid-edge}$ is defined to be a point equidistant from two adjacent vertices on a lattice. A \emph{domain} $\Omega$ is a simply connected collection of mid-edges on the half-plane honeycomb lattice. The set of vertices of the half-plane honeycomb lattice is denoted $V(\Omega)$. Those mid-edges of $\Omega$ which are adjacent to only one vertex in $V(\Omega)$ form $\partial\Omega$. 
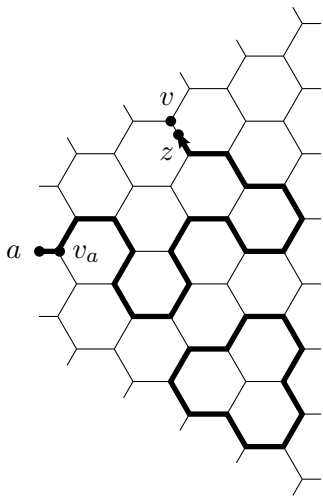
\begin{figure}
\centering
\begin{picture}(150,180)
\put(0,180){\includegraphics[scale=0.5, angle=270]{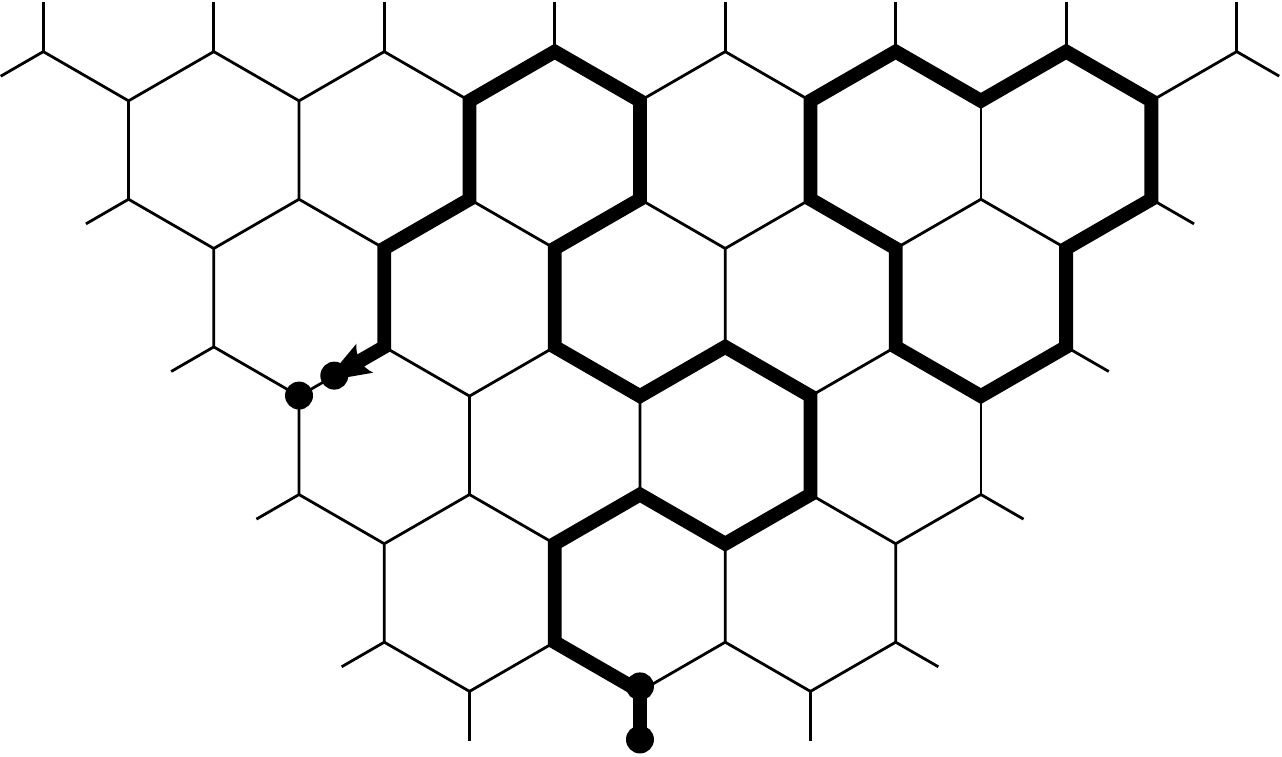}}
\put(-10,85){$a$}
\put(15,85){$v_a$}
\put(48,122){$z$}
\put(48,143){$v$}
\end{picture}
\caption{A configuration $\gamma$ on a finite domain. Point $a$ is a boundary mid-edge, point $z$ is another mid-edge, and $v_a$ and $v$ are corresponding vertices. The contribution of $\gamma$ to $F(z)$ is $\e^{-2\ii\sigma\pi/3}x^{30} n$.
} \label{fig:exampleF}
\end{figure}
Let $\gamma$ be a configuration on a domain $\Omega$ comprising a single self-avoiding walk and a number (possibly zero) of closed loops. We denote by $\ell(\gamma)$ the number of vertices occupied by $\gamma$ and $c(\gamma)$ the number of closed loops. Furthermore let $W(\gamma)$ be the winding angle of the self-avoiding walk component. Define the following observable.
\begin{definition}[Preholomorphic observable]
\label{def:Fdef}
\mbox{}\newline
\begin{itemize}
\item For $a\in\partial\Omega, z\in\Omega$, set
\be
F(\Omega,z;x,n,\sigma):=F(z) = \sum_{\gamma: a\rightarrow z} \e^{-\ii \sigma W(\gamma)} x^{\ell(\gamma)} n^{c(\gamma)},
\label{eq:Fmidedge}
\ee
where the sum is over all configurations $\gamma$ for which the SAW component runs from the mid-edge $a$ to a mid-edge $z$ (we say that $\gamma$ ends at $z$).
\item Let $F(p;v)$ only include configurations where there is a walk terminating at the mid-edge $p$ adjacent to the vertex $v$ and the other two mid-edges adjacent to $v$ are not occupied by a loop segment. For $v_a, v\in V(\Omega)$ and $p, q$ and $r$ mid-edges adjacent to $v$, set
\begin{align}
\overline{F}(V(\Omega),v;x,n,\sigma):=\overline{F}(v) = (p-v)F(p;v)+(q-v)F(q;v)+(r-v)F(r;v),
%= -\sum_{\gamma:v_a\rightarrow v} \e^{\ii (1-\sigma) W(\gamma)} x^{\ell(\gamma)} n^{c(\gamma)},
\label{eq:Fvertex}
\end{align}
Since this is a function involving walks that terminate at mid-edges adjacent to the vertex $v$ we consider this as a function defined at the vertices of the lattice.
%
%where $\tilde{\sigma}=1-\sigma$. The sum in the second line of \eqref{eq:Fvertex} is over all configurations $\gamma$ for which the SAW component runs from vertex $v_a$ to a vertex $v$.
\end{itemize}
See Fig.~\ref{fig:exampleF} for an example:
\end{definition}
Smirnov \cite{Smirnov10} proves the following: 
\begin{lemma}[Smirnov]
\label{lem:Slemma4}
For $n\in[-2,2]$, set $n=2\cos\phi$ with $\phi\in[0,\pi]$. Then for
\begin{align}
\sigma &= \frac{\pi-3\phi}{4\pi},\qquad x^{-1}=\x^{-1}=2\cos\left(\frac{\pi+\phi}{4}\right) = \sqrt{2-\sqrt{2-n}},\qquad\text{or}
\label{eq:Slemma_dense}\\
\sigma &= \frac{\pi+3\phi}{4\pi},\qquad x^{-1}=\x^{-1}=2\cos\left(\frac{\pi-\phi}{4}\right) = \sqrt{2+\sqrt{2-n}},
\label{eq:Slemma_dilute}
\end{align}
the observable $F$ satisfies the following relation for every vertex $v\in V(\Omega)$:
\begin{equation} \label{eqn:localidentity}
(p-v)F(p) + (q-v)F(q) + (r-v)F(r)=0,
\end{equation}
where $p,q,r$ are the mid-edges adjacent to $v$.
\end{lemma}

The first equation in Lemma~\ref{lem:Slemma4} corresponds to the larger of the two critical values of the step weight $x$ and thus describes the so-called dense regime as configurations with many loops are favoured. The second equation corresponds to the line of critical points separating the dense and dilute phases \cite{N82}. Eqn. (\ref{eqn:localidentity}) can be interpreted as the vanishing of a discrete contour integral, hence the name preholomorphic observable for $F(z)$.
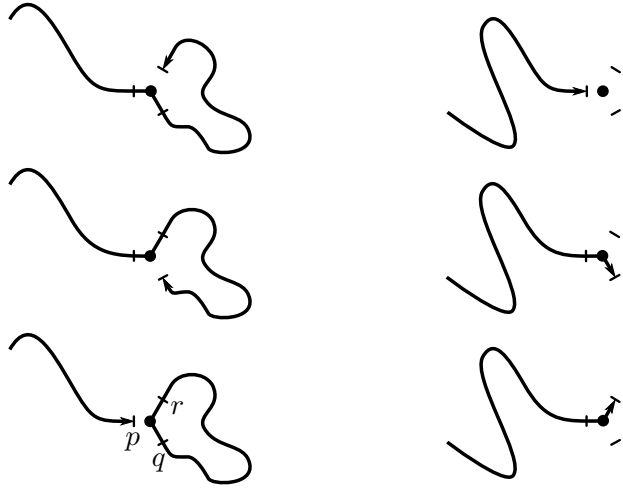
\begin{figure}
\centering
\begin{picture}(250,180)
\put(0,180){\includegraphics[scale=1.1, angle=270]{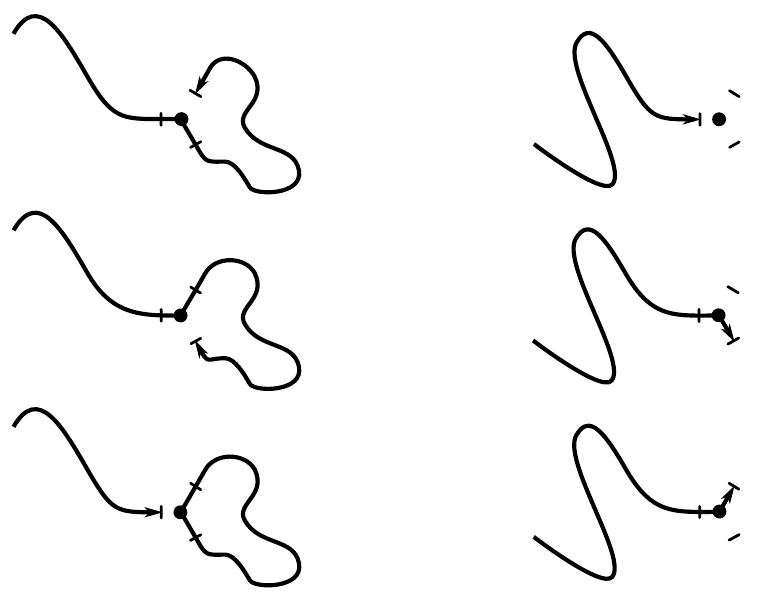}}
\put(48,8){$p$}
\put(65, 20){$r$}
\put(58, 0){$q$}
\end{picture}
\caption{The two types of configurations which end at mid-edges $p,q,r$ adjacent to vertex $v$. The first type, on the left, involves configurations which visit all three mid-edges. On the right are those configurations where the self-avoiding walk visits at most two mid-edges.} 
\label{fig:identity_groups}
\end{figure}
%\begin{figure}
%\centering
%\includegraphics[scale=1.1, angle=270]{identity_groups}
%\caption{The two types of configurations which end at mid-edges $p,q,r$ adjacent to vertex $v$. The first type, on the left, involves configurations which visit all three mid-edges. %On the right are those configurations where the self-avoiding walk visits at most two mid-edges.}
%\label{fig:identity_groups}
%\end{figure}
\begin{proof}
Consider a vertex $v$ adjacent to a mid-edge $p$. The two other adjacent mid-edges we refer to as $q$ and $r$ and are labelled as shown in Fig.~\ref{fig:identity_groups}. For a self-avoiding walk entering the vertex $v$ from the mid-edge $p$ and terminating at either $p, q$ or $r$ there are two disjoint sets of configurations to consider, each corresponding to a different external connectivity of the remaining mid-edges $q$ and $r$. These are also shown in Fig. (\ref{fig:identity_groups}). Since the two sets of configurations are disjoint we can consider the identity (\ref{eqn:localidentity}) for each case separately.
%There are two possible cases to consider, each corresponding to a different external connectivity of the mid-edges adjacent to the vertex. 
In the following, we define $\lambda=\e^{-\ii\sigma\pi/3}$ (the weight accrued by a walk for each left turn) and $j=\e^{2\ii\pi/3}$ (the value of $p-v$ when mid-edge $p$ is to the north-west of its adjacent vertex $v$).
\begin{enumerate}
\item
In the first case, we consider all configurations where mid-edges $p$ and $q$ are connected. There are three ways for this to occur: two with the self-avoiding walk visiting all three sites, and one with a closed loop running through $v$. Furthermore, we define $F_L(P;v)$ to be the contribution to $F(p)$ involving only configurations where the walk ends at the point $p$, adjacent to the vertex $v$ and where the two other mid-edges adjacent to $v$ are occupied by a closed loop. Requiring \eqref{eqn:localidentity} to hold then implies
\be
(p-v)F_L(p;v)+(q-v)\frac{1}{n}\bar{\lambda}^4F_L(p;v)+(r-v)\frac{1}{n}\lambda^4F_L(p;v)=0.
\label{eqn:localidentity2}
\ee
The factor of $\frac{1}{n}$ arises from the absence of a closed loop and the complex phase factors arise from the additional winding: the loop makes an additional four left turns to arrive at $q$ and four right turns to arrive at $r$. Substituting these into \eqref{eqn:localidentity2} we find
\be
\frac{1}{n} (p-v)F_L(p;v)(-n-\bar{\lambda}^4 j- \lambda^4\bar{j} )=0, \nonumber
\ee
where we have used that $q-v=j(p-v)$ and $r-v=\bar{j}(p-v)$. Since the choice of $v$ and $p$ was arbitrary this implies
\be
n+\lambda^4 \bar{j} +\bar{\lambda}^{4}j=0. \nonumber
\ee
Finally, using the parameterisation of $n$ in terms of $\phi$ and solving for $\sigma$ we obtain
\begin{align}
\sigma &= \frac{\pi-3\phi}{4\pi} \qquad \text{for } \lambda^4 \bar{j} = -\e^{\ii\phi},
\label{eq:sigmadense}\\
\sigma &= \frac{\pi+3\phi}{4\pi} \qquad \text{for } \lambda^4 \bar{j} = -\e^{-\ii\phi},
\label{eq:sigmadilute}
\end{align}
\item
In the second case only one or two mid-edges are occupied in the configuration and mid-edges $q$ and $r$ are not connected.
Recalling the definition of $F(p; v)$ in (\ref{def:Fdef}) and Eqn. \eqref{eqn:localidentity} we have
\be
(p-v)F(p; v)+(q-v)x\bar{\lambda} F(p; v)+(r-v)x\lambda F(p; v)=0,
\ee
which simplifies to
\be
F(p; v)(-1-\x\bar{\lambda}j-\x\lambda\bar{j})=0.
\ee
Again, since this equation holds for arbitrary $v$ we obtain
\be
1+\x\lambda\bar{j}+\x\bar{\lambda}j=0,
\ee
which leads to
\be
\x^{-1}=2\cos\left(\frac\pi3(\sigma-1)\right).
\ee
\end{enumerate}
The two possible values of $\sigma$ give rise to the corresponding two values for $\x$.
\end{proof}

%%%%%%%%%%%%%%%%%%%%%%%%%%%%%%%%%%%%%%%%%%%%%%%%%
\subsection{Off-critical deformation}
%%%%%%%%%%%%%%%%%%%%%%%%%%%%%%%%%%%%%%%%%%%%%%%%%
First we evaluate the discrete divergence of the second set of configurations in Fig.~\ref{fig:exampleF} for general $x$, below the critical value. This gives
\begin{lemma}[Massive preholomorphicity identity]
\label{lem:massiveCR}
For a given vertex $v$ with mid-edges $p$, $q$ and $r,$ and $x$ below the critical value $\x$, the parafermionic observable $F(z)$ satisfies
\begin{align}
\label{eqn:massiveCR}
(p-v)F(p)+(q-v)F(q)+(r-v)F(r)=(1-\frac{x}{\x}) \overline{F}(v),
\end{align}
where $F(z)$ and $\overline{F}(v)$ are defined in Definition \ref{def:Fdef}.
\end{lemma}
We use the term massive preholomorphic as \eqref{eqn:massiveCR} is of a similar form to that described in \cite{MS09} and \cite{Smirnov10}.
\begin{proof}
Similar to Lemma~\ref{lem:Slemma4} the proof splits into two parts. The first part, concerning cancellations of contributions coming from walks depicted in the left-hand side of Fig. \ref{fig:identity_groups}, is completely analogous, and as before fixes the value of $\sigma$. The difference is that now we relax the requirement that the contribution from the second set of configurations (shown on the right in Fig.~\ref{fig:identity_groups}) to the discrete contour integral vanishes. Consequently,  $x$ is no longer fixed to the critical value.

Consider a vertex $v$ with mid-edges labelled $p$, $q$ and $r$ in a counter-clockwise fashion. There are three disjoint sets of configurations, depending on which of the three mid-edges $p$, $q$ or $r$ the walk enters from. These are shown in Fig. \ref{fig:pqrvertex}. Recall that we denote by $F(p;v)$ the contributions to $F(p)$ that only include configurations where there is a walk terminating at the mid-edge $p$ adjacent to the vertex $v$ and where the two other mid-edges adjacent to $v$ are unoccupied. The contribution to the left-hand side of \eqref{eqn:massiveCR} from walks entering the vertex from $p$ is the sum of three terms
\begin{eqnarray}
\label{eqn:venterp}
&&(p-v)F(p;v)+(q-v)x\,\e^{\pi\sigma\ii/3}F(p;v)+(r-v)x\,\e^{-\pi\sigma\ii/3}F(p;v).
\end{eqnarray}
The first term is simply from walks that enter and terminate at $p$. The second term is from those walks that enter from $p$, make a right turn and terminate at $q$. The final term is  from walks that enter at $p$ and make a left turn to terminate at $r$. The last two terms acquire an additional weight $x$ from the extra step and a phase factor from the turn. We can simplify \eqref{eqn:venterp} to obtain
\begin{eqnarray*}
&=& (p-v)F(p;v)+(p-v)xj\bar{\lambda}F(p;v)+(p-v)x\lambda \bar{j}F(p;v) \\
&=& (p-v)F(p;v)(1-xj\bar{\lambda} -x\bar{j}\lambda)\\
&=& (p-v)F(p;v)(1-\frac{x}{\x}),
\end{eqnarray*}
where in the first line we have used that 
\[
q-v=j(p-v),\qquad r-v=\bar{j}(p-v).
%\label{eq:vectors}
\]
For walks entering from mid-edges $q$ and $r$ similar calculations give contributions
\[
(q-v)F(q;v)(1-\frac{x}{\x}) \text{ and }
(r-v)F(r;v)(1-\frac{x}{\x}).
\]
Adding the three contributions together and using Definition~\ref{def:Fdef} gives the right-hand side of Eqn. \eqref{eqn:massiveCR}. 
\end{proof}
Using the above lemma we can now derive the following relationship between generating functions. \\
\begin{lemma}[Off-critical generating function identity]
\label{lem:offcriticalDCS}
\begin{align}
\sum_{\gamma : a \to z\in\partial \Omega\backslash\{a\}} e^{\ii \tilde{\sigma} W(\gamma)} x^{|\gamma|}n^{c(\gamma)}  + (1- x/x_c)\sum_{\gamma : a \to z\in\Omega\backslash \partial \Omega} e^{\ii \tilde{\sigma} W(\gamma)} x^{|\gamma|}n^{c(\gamma)}  = C_{\Omega}(x),
\end{align}
where
\be
C_{\Omega}(x) = \sum_ {\gamma : a \to a} x^{|\gamma|}n^{c(\gamma)},
\ee
is the usual generating function of the honeycomb lattice O($n$) model, i.e. closed loops without the SAW component, and $\tilde{\sigma} = 1-\sigma$.
\end{lemma}
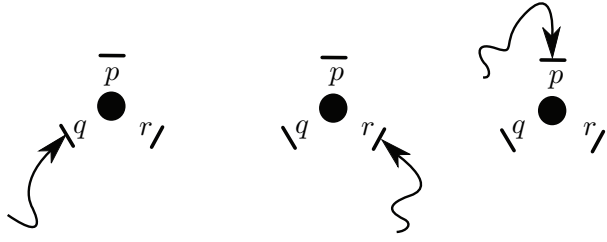
\begin{figure}
\centering
\begin{picture}(400,100)
\put(85,0){\includegraphics[scale=0.75, angle=0]{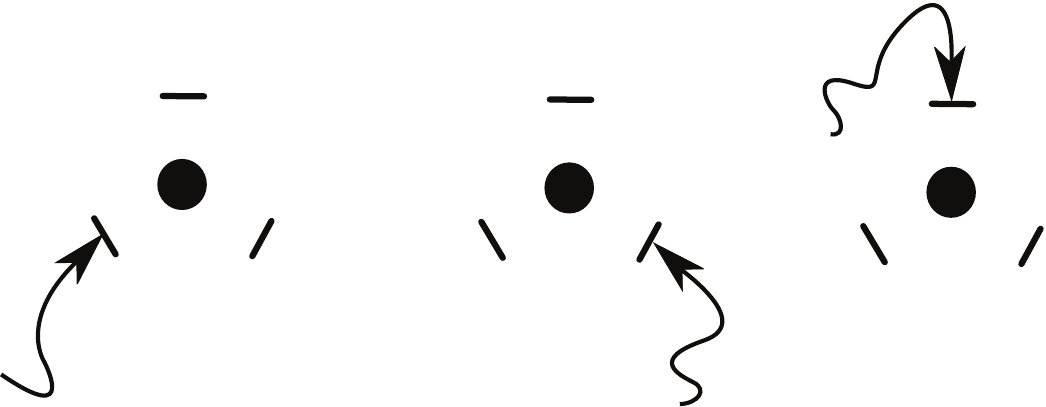}}
\put(122,58){$p$}
\put(207,58){$p$}
\put(290,57){$p$}
\put(110,38){$q$}
\put(194,38){$q$}
\put(276,38){$q$}
\put(135,37){$r$}
\put(219,37){$r$}
\put(303,37){$r$}
\end{picture}
\caption{The three possible ways for a walk to enter a given vertex via each of the three mid-edges, $p$, $q$ and $r$. The discrete divergence is evaluated for all three cases in order to derive the off-critical, or `massive' preholomorphicity condition.}
\label{fig:pqrvertex}
\end{figure}
\begin{proof}
We begin by summing Eqn. \eqref{eqn:massiveCR} over all the vertices of the lattice $\Omega$. The contribution to the left-hand side of \eqref{eqn:massiveCR} from those mid-edges that are in the bulk cancels, since each bulk mid-edge is summed over twice but with opposite signs. This leaves only the boundary mid-edges contributing to the left-hand side which can be written as
\be
\sum_{\gamma : a \to z \in\partial \Omega} \e^{ -\ii \sigma W(\gamma)} x^{|\gamma|} n^{c(\gamma)}\e^{\ii\phi(\gamma)}, \nonumber
\ee
where $\e^{\ii\phi(\gamma)}$ is the complex number that describes the direction from the boundary vertex to the boundary mid-edge. It is easy to check that this equals $\e^{\ii W(\gamma)}$ for all walks terminating on boundary mid-edges other than the starting mid-edge $a$ and is $-1$ if the walk terminates at $a$ (which we call a $\emph{zero-length walk}$) . Using $\tilde{\sigma}=1-\sigma$ we then have
\be
\label{eqn:lhssum}
\sum_{\gamma : a \to z\in\partial \Omega\backslash\{a\}} \e^{\ii \tilde{\sigma} W(\gamma)} x^{|\gamma|} n^{c(\gamma)}-\sum_{\gamma : a \to a} x^{|\gamma|} n^{c(\gamma)}.
\ee
This first term arises from all configurations where the walk terminates on a boundary mid-edge other than the starting mid-edge $a$. The second is from all configurations with a zero-length walk, that is one that terminates at $a$. Note that we define the winding angle of a zero-length walk to be $0$.

As for the right-hand side of Eqn. \eqref{eqn:massiveCR}, using Definition \ref{def:Fdef} this can be written as
\be
\label{eqn:rhssum}
\left (1-\frac{x}{x_{\rm c}} \right ) \sum_{\gamma : a \to z \in\Omega\backslash \partial \Omega} \left [ F(z;v_1(z)) (z-v_1(z))+ F(z;v_2(z)) (z-v_2(z)) \right ].
\ee
This is because for a given end point $z$, a walk can be heading towards one of two possible vertices which we call $v_1$ and $v_2$, the labelling being unimportant.  This is illustrated in Fig. \ref{fig:2vertex}. Equating \eqref{eqn:lhssum} and \eqref{eqn:rhssum} we have
\begin{align}
%& & j\sum_{\gamma : a \to z \in \epsilon} F(z)+\bar{j}\sum_{\gamma : a \to z \in \bar{\epsilon}} F(z)-\sum_{\gamma : a \to z \in \alpha} F(z)+\sum_{\gamma : a \to z\in\beta}F(z) \\
&\sum_{\gamma : a \to z\in \partial \Omega\backslash \{a\}}\e^{ \ii \tilde{\sigma} W(\gamma)} x^{|\gamma|}n^{c(\gamma)}-\sum_{\gamma : a \to a}x^{|\gamma|}n^{c(\gamma)} \nonumber\\
= &\left (1-\frac{x}{x_{\rm c}} \right ) \sum_{\gamma : a \to z \in\Omega\backslash \partial \Omega} \left ( F(z;v_1) (z-v_1(z))+ F(z;v_2) (z-v_2(z)) \right ) 
\label{eqn:identity1}
\end{align}
%
%where $\epsilon$, $\bar{\epsilon}$, $\alpha$ and $\beta$ refer respectively to the set of mid-edges on the upper, lower, left and right boundaries of the domain $\Omega$. \\ \\
%
Using $\sigma=1-\tilde{\sigma}$ and the definition of $F(z;v)$ the summation on the right-hand side becomes 
\[
\e^{ \ii \phi } \left ( \sum_{\gamma : a \to z \to v_1} x^{|\gamma|}n^{c(\gamma)}\e^{ \ii \tilde{\sigma} W(\gamma)}\e^{-\ii W(\gamma)} - \sum_{\gamma : a \to z \to v_2} x^{|\gamma|}n^{c(\gamma)}e^{\ii \tilde{\sigma} W(\gamma)} \e^{-\ii W(\gamma)}  \right ), 
\] \\
where $e^{\ii\phi}$ is the unit vector from $v_1$ to $z$, which is the negative of the unit vector from $v_2$ to $z$.
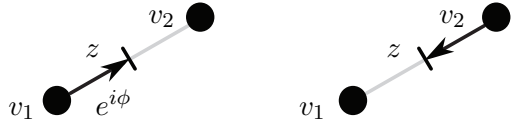
\begin{figure}
\centering
\begin{picture}(250,75)
\put(50,0){\includegraphics[scale=0.75, angle=0]{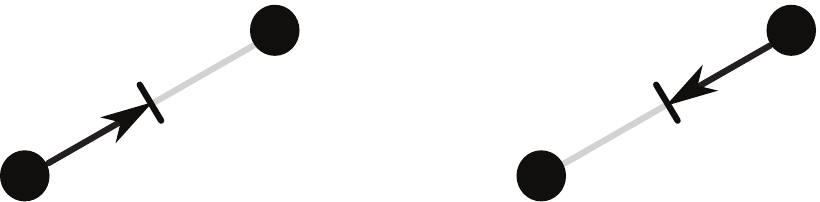}}
\put(37,0){$v_1$}
\put(90,35){$v_2$}
\put(66, 22){$z$}
\put(147,0){$v_1$}
\put(200,35){$v_2$}
\put(180, 22){$z$}
\put(70,0){$e^{i\phi}$}
\end{picture}
\caption{A walk terminating at the mid-edge $z$. The mid-edge lies between two vertices $v_1$ and $v_2$ and the unit vector from $v_1$ to $z$ is given by $e^{\ii \phi}$. The labelling of the vertices is arbitrary.} 
\label{fig:2vertex}
\end{figure}

A walk that terminates at $z$ and moves in the direction of vertex $v_2$ has winding $W(\gamma_2)=2\pi m' + \phi$ while a walk heading in the direction of the vertex $v_1$ and terminating at $z$ has winding $W(\gamma_1)=(2m + 1)\pi + \phi$ for some $m, m' \in\mathbb{Z}$. In each case the angle $\phi$ from the unit vector is cancelled by the $\phi$ appearing in the winding angle term $e^{-\ii W(\gamma)}$ and this leaves
\be
\label{exp:rhsidentity1}
-\sum_{\gamma : a \to z \in \Omega\backslash \partial \Omega} x^{|\gamma|}n^{c(\gamma)}\e^{ \ii \tilde{\sigma} W(\gamma)}. 
\ee

The left-hand side \eqref{eqn:lhssum} is a sum of walks to the boundary and walks of length zero, which is equal to $C_{\Omega}(x)$. Substituting expression (\ref{exp:rhsidentity1}) into Eqn. (\ref{eqn:identity1}) and rearranging we obtain 
\begin{equation} \label{eqn:keyidentity}
\sum_{\gamma : a \to z\in \partial \Omega\backslash \{a \}} \e^{\ii\tilde{\sigma} W(\gamma)} x^{|\gamma|}n^{c(\gamma)}  + (1- x/x_c)\sum_{\gamma : a \to z\in \Omega\backslash \partial \Omega} \e^{\ii\tilde{\sigma} W(\gamma)} x^{|\gamma|}n^{c(\gamma)}  = C_{\Omega}(x)
\end{equation}
\end{proof}
%%%%%%%%%%%%%%%%%%%%%%%%%%%%%%%%%%%%%%%%%%%%%%%%%
\section{Winding angle}
%%%%%%%%%%%%%%%%%%%%%%%%%%%%%%%%%%%%%%%%%%%%%%%%%
\subsection{Generating function definitions}
\begin{figure}[t]
\begin{center}
\begin{picture}(140,190)
\put(0,180){\includegraphics[height=120pt, angle=270]{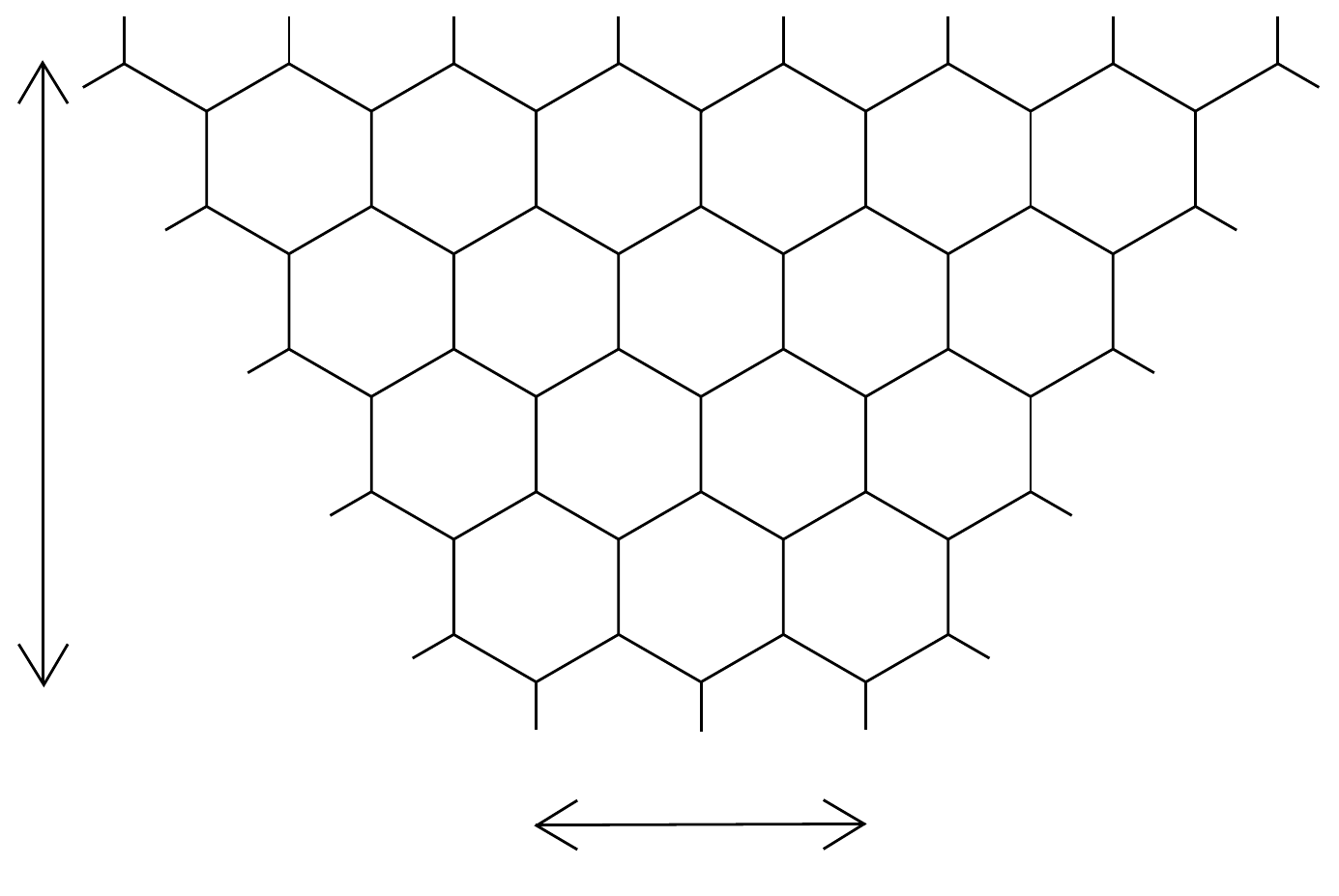}}
\put(15,77){$\alpha$}
\put(123,76){$\beta$}
\put(55,143){$\varepsilon$}
\put(55,10){$\bar{\varepsilon}$}
\put(15,89){$a$}
\put(-11,77){$2L$}
\put(69,182){$T$}
\end{picture}
\end{center}
\caption{Finite patch $S_{5,1}$ of the hexagonal lattice. The SAW component of a loop configuration starts on the central mid-edge of the left boundary (shown as $a$).}
\label{fig:S_boundary}
\end{figure}
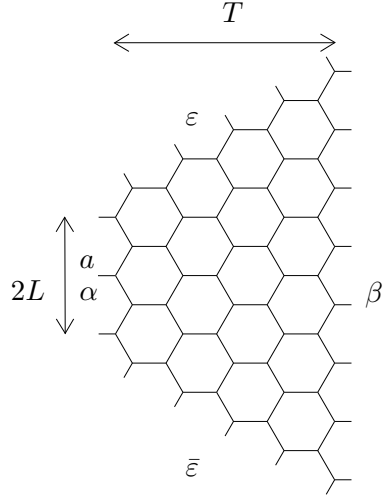

Let us now restrict to a particular trapezoidal domain $\Omega=S_{T,L}$ of width $T$ and left-height $2L$, see Fig.~\ref{fig:S_boundary}.
The winding angle distribution function can be calculated directly from the off-critical generating function identity (\ref{eqn:keyidentity}). We remind the reader that $\gamma$ describes a walk along with a configuration of loops. For convenience we use the terms $\emph{generating function of walks}$ and $\emph{configuration of walks}$ but it should be understood that these include configurations of closed loops as well. We define the following generating function
\[
G_{\theta,\Omega}(x)=\sum_{\substack{\gamma : a \to z\in\Omega\backslash \partial \Omega\\W(\gamma)=\theta}}x^{|\gamma|}n^{c(\gamma)}.
\]
$G_{\theta,\Omega}(x)$ contains only those contributions to $G_\Omega(x) = \sum_\theta G_{\theta,\Omega}(x)$ where the walk has winding angle $\theta$. We also define
\[
H_{\Omega}(x)=\sum_{\gamma : a \to z\in\partial \Omega\backslash \{a \} }\e^{\ii\tilde{\sigma} W(\gamma)}x^{|\gamma|}n^{c(\gamma)},
\]
which is the generating function describing walks that terminate on the boundary of the domain, and thus have a winding angle associated to that boundary.
Using this notation (\ref{eqn:keyidentity}) becomes
\[
H_{\Omega}(x)  + (1- x/x_c)\sum_{\theta} \e^{\ii\tilde{\sigma} \theta}G_{\theta,\Omega}(x) = C_{\Omega}(x).
\]
Now let $H_{\Omega}^{*}(x)$ and $G_{\theta,\Omega}^{*}(x)$ be $H_{\Omega}(x)/C_{\Omega}(x)$ and $G_{\theta,\Omega}(x)/C_{\Omega}(x)$ respectively. For $x<x_{\rm c}$ define $H^{*}(x)$ and $G_{\theta}^{*}(x)$ to be $H_{\Omega}^{*}(x)$ and $G_{\theta,\Omega}^{*}(x)$ respectively in the limit as $\Omega$ approaches the half plane. Assuming that $\x$ is the location of the critical point\footnote{For SAWs ($n=0$) it was proved in \cite{DC-S10} that $x_{\rm c}$ is indeed the critical point. Likewise for the $n=1$ case \cite{H50} and the $n=-2$ case \cite{BT73} this is rigorously known.}, as we will do in the next section, the limits of $H_{\Omega}^{*}(x)$ and $G_{\theta,\Omega}^{*}(x)$ exist for $x<\x$ by definition of the critical point.

Moreover, since $H_{\Omega}^{*}(x)$ converges, the contributions from configurations whose walk ends at the top/bottom or right-hand boundary are tail terms of a converging series and hence vanish as $\Omega$ approaches the half-plane. Thus $H^{*}(x)$ only contains walks starting and ending at the surface $\alpha$, which is the only surface remaining in the domain. In the limit as the strip width becomes infinite we thus obtain

\begin{equation}\label{eqn:f*g*}
H^{*}(x)  + (1- x/x_c)\sum_{\theta} \e^{\ii\tilde{\sigma} \theta}G^{*}_{\theta}(x) = 1.
\end{equation}

\subsection{Susceptibilities and critical exponents}
The first term $H^{*}(x)$ is (up to a normalisation) the generating function of walks that start and end at the $\alpha$ surface (and have an additional half-step to the left of their starting and ending vertices). This generating function is usually denoted in the literature as $\chi_{11}(x)$ \cite{BH72}. One conventionally writes 
\[
\chi_{11}(x) \sim d_0(x)+d_1(x)(1-x/x_{\rm c})^{-\gamma_{11}},\qquad x\lesssim x_{\rm c},
\]
where $d_0$ and $d_1$ are analytic at $x=x_{\rm c}$ and by $\sim$ we mean that the ratio of the left and right sides approaches $1$ as $x$ approaches the critical point. In doing this we are assuming the existence of the exponent $\gamma_{11}$ as well as that the value of the critical exponent is given by $x_{\rm c}$ as defined in \eqref{eq:Slemma_dilute}.

Similarly, the generating function of walks that start at the surface and finish anywhere inside the domain, is usually denoted in the literature as $\chi_{1}(x).$ One conventionally writes 
\[
\chi_{1}(x) \sim c(x)(1-x/x_{\rm c})^{-\gamma_{1}}, \qquad x\lesssim x_{\rm c},
\]
with $c(x_c)$ analytic. This assumes the existence of the exponent $\gamma_{1}$ and the value of the critical point, but no other assumption is made. We will see later that this assumption for example is not valid at $n=2$. 

With the assumption as to the existence of the exponents for the susceptibilities we have
\[
H^{*}(x)\propto \chi_{11}(x) \sim 1+const \times(1-x/x_{\rm c})^{-\gamma_{11}},
\]
and
\be
\sum_{\theta}G_{\theta}^{*}(x) \propto \chi_1(x) \sim const\times (1-x/x_{\rm c})^{-\gamma_{1}},
\label{eq:Gasymp}
\ee
Using Eqn. (\ref{eqn:f*g*}) we also obtain
\begin{equation}\label{eqn:g*}
\sum_{\theta} e^{\tilde{\sigma} i \theta}G^{*}_{\theta}(x)=(1-H^{*}(x))(1- x/x_c)^{-1}\sim const\times (1-x/x_{c})^{-\gamma_{11}-1}.
\end{equation}

\subsection{Asymptotic winding angle distribution}
We first state the following result regarding the asymptotic expansion of certain singular functions \cite{FS09}
\begin{thm}
\label{thm:singgam}
Let $\alpha$ be an arbitrary complex number in $\mathbb{C}\backslash\mathbb{Z}_{\leq 0}$. The coefficient of $z^n$ in
\[
f(z)=(1-z)^{-\alpha}
\]
admits for large $n$ a complete asymptotic expansion in descending powers of $n$,
\[
[z^n]f(z) \sim \frac{n^{\alpha-1}}{\Gamma(\alpha)}\left(1+O \left ( \frac{1}{n} \right ) \right ).
\]
\end{thm}
This allows one to show the following:
\begin{lemma}
\label{lem:GFasymp}
Let $G(x)=\sum_{j\ge 0} g_j x^j$. If $G(x)\sim A (1-x/x_{\rm c})^{-\eta}$ for $x\lesssim x_{\rm c}$ and some constant $A$, then
\[
g_j \sim A x_{\rm c}^{-j} j^{\eta-1}/\Gamma(\eta).
\]  
\end{lemma}

We denote the number of walks of length $j$ with winding angle exactly $\theta$ by $a_{\theta}(j)$, so that we can write
\[
G^{*}_{\theta}(x)=\sum_{j=0}^{\infty}a_{\theta}(j)x^{j}.
\]
Summing over $\theta$ and using \eqref{eq:Gasymp} and Lemma~\ref{lem:GFasymp}, the total number of walks of length $j$ behaves like
\[
\sum_{\theta}a_{\theta}(j)\sim const\times x_{\rm c}^{-j}j^{\gamma_{1}-1}.
\]
Similarly, from (\ref{eqn:g*}) we have 
\[
\sum_{\theta} \e^{\ii\tilde{\sigma} \theta}a_{\theta}(j)\sim const\times x_{\rm c}^{-j}j^{\gamma_{11}}.
\]

\begin{definition}
The probability density function $P(\theta,j)$ for the winding angle of walks of length $j$ is the fraction of walks of length $j$ with winding angle $\theta$: 
\[
P(\theta,j) = \frac{a_{\theta}(j)}{\sum_{\theta}a_{\theta}(j)}.
\]
\end{definition}

From the reasoning above, the following result follows immediately.
\begin{prop}
\label{prop:Psurf}
Let $\tilde{\sigma}=1-\sigma$ where $\sigma$ is given by \eqref{eq:sigmadilute}. Then
\[
\sum_{\theta} \e^{\ii\tilde{\sigma} \theta}P(\theta,j)\sim const\times j^{\gamma_{11}-\gamma_{1}+1}.
\]
\end{prop}
That is to say, the probability density function is characterised by an exponent that can be expressed solely in terms of the half-plane exponents $\gamma_1$ and $\gamma_{11}.$
%%%%%%%%%%%%%%%%%%%%%%%%%%%%%%%%%
\subsection{Winding angle in a wedge}\label{ssec:wedge}
%%%%%%%%%%%%%%%%%%%%%%%%%%%%%%%%%

Rather than taking $L\rightarrow\infty$ as in the previous sections, we now set $L=0$. In this case the trapezoidal domain $S_{T,0}$ reduces to a wedge with wedge angle $\alpha=\pi/3$. Using similar arguments as before, we take the limit $T\rightarrow\infty$, giving
\begin{equation}\label{eqn:f*g*wedge}
H^{*}_\alpha(x)  + (1- x/x_{\rm c})\sum_{\theta} \e^{\ii\tilde{\sigma} \theta}G^{*}_{\theta,\alpha}(x) = 1,
\end{equation}
where now $H^{*}_\alpha(x)$ is (again up to a normalisation factor) the generating function of walks that start at $a$ and end at the $\varepsilon$ or $\bar{\varepsilon}$ surface of the wedge with angle $\alpha=\pi/3$. This generating function is usually denoted in the literature $\chi_{21}(x,\alpha)$ \cite{GT84}. It is known that self-avoiding walks ($n=0$) in a wedge have the same connective constant as those in the plane, for arbitrary wedge angle \cite{HW85}. This is also known to be true in the Ising ($n=1$) case \cite{BPP84}, and for the O$(-2)$ model, the Gaussian case \cite{BT73}. We assume that this holds also for all $n\in[-2,2]$. We thus write 
\[
\chi_{21}(x,\pi/3) \sim \tilde{d}_0(x)+\tilde{d}_1(x)(1-x/x_{\rm c})^{-\gamma_{21}(\pi/3)},
\]
where $\tilde{d}_0$ and $\tilde{d}_1$ are analytic near $x=x_{\rm c}$. This assumes the existence of the exponent $\gamma_{21}(\alpha)$ which in general depends on the wedge angle $\alpha.$ Similarly, the generating function of walks that start at $a$ and finish anywhere inside the domain, is usually denoted in the literature as $\chi_{2}(x,\alpha).$ Assuming the existence of the relevant critical exponent $\gamma_{2}(\alpha)$, we write 
\[
\chi_{2}(x,\pi/3) \sim \tilde{c}(x)(1-x/x_{\rm c})^{-\gamma_{2}(\pi/3)},
\]
with $\tilde{c}(x)$ analytic near $x = x_c.$  

Denote the probability density function for the winding angle of walks of length $j$ in a wedge with angle $\alpha$ by $P_\alpha(\theta,j)$. Using exactly the same reasoning as in the previous section, we find that
\be
\sum_{\theta} \e^{\ii\tilde{\sigma} \theta}P_{\alpha=\pi/3}(\theta,j)\sim const\times j^{\gamma_{21}(\pi/3)-\gamma_{2}(\pi/3)+1}.
\ee
In fact, one can approximate every wedge angle by an appropriate stacking of additional boundary vertices and mid-edges. Using the same reasoning as before we can therefore generalise to arbitrary wedge angle and find:
\be
\sum_{\theta} \e^{\ii\tilde{\sigma} \theta}P_{\alpha}(\theta,j)\sim const\times j^{\gamma_{21}(\alpha)-\gamma_{2}(\alpha)+1}.
\label{eq:Pwedge}
\ee
This reduces to the previous result in the special case $\alpha = \pi.$ 

%%%%%%%%%%%%%%%%%%%%%%%%%%%%%%%%%%%
\subsection{Exponent inequalities}
%%%%%%%%%%%%%%%%%%%%%%%%%%%%%%%%%%%
Using the techniques employed above we can derive rigorous exponent bounds in the following way. Recall
\[
H^{*}(x)\propto \chi_{11}(x) \sim 1+const \times(1-x/x_{\rm c})^{-\gamma_{11}},
\]
and
\[
\sum_{\theta}G_{\theta}^{*}(x) \propto \chi_1(x) \sim const\times (1-x/x_{\rm c})^{-\gamma_{1}}.
\]
Since $$\sum_{\theta} \e^{\ii\tilde{\sigma} \theta}G_{\theta}^{*}(x) \le \sum_{\theta}G_{\theta}^{*}(x) ,$$ it follows from
\[
H^{*}(x)  + (1- x/x_{\rm c})\sum_{\theta} \e^{\ii\tilde{\sigma} \theta}G^{*}_{\theta}(x) = 1,
\]
that
\be 
\gamma_1 -\gamma_{11} \ge 1.
\label{ineq}
\ee
The only assumptions here are the existence of the critical exponents and the assumption that the critical point is at $x = x_c,$ as given by Lemma \ref{lem:Slemma4} (except for for $n=0, \,\, 1$ and $-2,$ where it is known rigorously \cite{DC-S10,H50, BT73}). To see how strong the inequality is, one must substitute the conjectured exponent values. For the O$(n=-2)$ model, the inequality is an equality. As $n$ increases, the bound gets progressively weaker. For the O$(n=-1)$ model, the l.h.s. of \eqref{ineq} is $67/64,$  for the O$(n=0)$ model, the l.h.s. is $73/64,$ while for the O$(n=1)$ model it is $11/8$. The exponents do not exist for the O$(2)$ model.

As we have shown that $\gamma_2(\alpha) -\gamma_{21}(\alpha)$ is independent of $\alpha,$ it follows that the above inequality can be written more generally as 
\be 
\gamma_2(\alpha) -\gamma_{21}(\alpha) \ge 1.
\label{ineq2}
\ee

%%%%%%%%%%%%%%%%%%%%%%%%%%%%%%%%%%%%%%%
\section{Conjectures}
%%%%%%%%%%%%%%%%%%%%%%%%%%%%%%%%%%%%%%%
\subsection{Winding angle distribution from conformal field theory}
%%%%%%%%%%%%%%%%%%%%%%%%%%%%%%%%%%%%%%%
Duplantier and Saleur \cite{DS88} conjectured the winding angle distribution function for the general O($n$) model on a cylinder. We use  the parametrisation $n=-2\cos(4\pi/\kappa)$ whereas Duplantier and Saleur use the symbol $g$ which is related to $\kappa$ by $g=4/\kappa$ ($\kappa=8/3$ for SAWs). The parafermionic spin $\sigma$ is related to $\kappa$ by $\sigma=\frac{3}{\kappa}-\frac{1}{2}$. They conjecture, from CFT and Coulomb gas arguments, that
\begin{equation}\label{DS}
P(x=\theta) \propto \exp\left(-\frac{\theta^2}{2\kappa\nu \log \ell}\right) ,\qquad\ell\rightarrow\infty, 
\end{equation}
where $\ell$ is the length of the walk. Here $\nu$ is the standard critical exponent relating the circumference of the cylinder to the length of the walk, $L\sim \ell^{\nu}$, and is given by
\[
\nu=\frac{1}{4-\kappa},
\]
where $\kappa = 2, \,\, 12/5, \,\, 8/3, \,\, 3, \,\, 4$ for $n = -2, \,\, -1,\,\, 0, \,\, 1, \,\, 2$ respectively. Hence $\nu$ is not defined for $n=2$. Note that the winding angle distribution \eqref{DS} is expressed entirely in terms of bulk critical exponents.  

We recall from the previous sections that
\begin{equation}\label{distrib}
\sum_{\theta} \e^{\ii\tilde{\sigma}\theta} P_{\alpha}(\theta,\ell) \approx \int_{-\infty}^\infty \e^{\ii\tilde{\sigma} \theta} P_\alpha(\theta,\ell) d\theta \propto \ell^{-\omega},
\end{equation}
where $\omega=-\gamma_{21}(\alpha)+\gamma_2(\alpha)-1$. The half plane corresponds to $\alpha=\pi$. 
Using (\ref{DS}) this is a straightforward integral, and gives
\[
\omega=\nu\kappa\tilde{\sigma}^2/2 = \frac{\kappa\tilde{\sigma}^2}{2(4-\kappa)}.
\]
%
%For SAW ($\kappa=8/3$) this gives $\omega=\tilde{\sigma}^2=9/64$. 
Hence we find
\be
\label{eq:surfbulk}
-\gamma_{21}(\alpha)+\gamma_2(\alpha)-1= \omega=\frac98\frac{(2-\kappa)^2}{\kappa(4-\kappa)}.
\ee
In particular we note that $\omega=-\gamma_{21}(\alpha)+\gamma_2(\alpha)-1$ is independent of the wedge angle $\alpha$. This is confirmed in the case of SAW ($n=0$) by the results of \cite{GT84}, and for the Ising case ($n=1$) by the results of \cite{BPP84}.

From the existing physics literature \cite{BM77,N82,C84} one can express the conjectured values of the half-plane exponents as follows: 
\begin{equation}\label{gam1}
\gamma_1=\frac{\kappa^2+12\kappa-12}{8\kappa(4-\kappa)},\qquad \gamma_{11}= -\frac{2(3-\kappa)}{\kappa(4-\kappa)},
\end{equation}
and thus
\[ 
-\gamma_{11}+\gamma_1 -1= \frac98\frac{(2-\kappa)^2}{\kappa(4-\kappa)},
\]
in perfect agreement with \eqref{eq:surfbulk} for $\alpha=\pi$.

\subsection{Wedge exponents}
The expected values of the wedge exponents have not previously been derived for general $n$. In \cite{BD08} this is done for the $n=0$ case with the assumption that $\mathrm{SLE}_{8/3}$ describes the scaling limit. However by extrapolating certain special cases we can provide conjectured values for these exponents for $n\in [-2,2)$. Following the notation of \cite{GT84}, we write the free energy of a $d$-dimensional wedge-shaped system as $$F=Vf_b + Af_s+Lf_e + \ldots,$$ where $V$ is the $d$-dimensional `volume' of the system, $A$ is the $(d-1)$-dimensional `area' of a surface, and $L$ is the $(d-2)$-dimensional `length' of the edge formed by the intersection of the two surfaces. In our case, $d=2,$ so the `volume' is an area, the `area' is the length of the boundary, and the `length' corresponds to the point at the apex of a wedge. Using the scaling hypothesis, the singular part of the  corresponding free energies can be written as
\begin{eqnarray}
f_b & \sim & t^{2-\alpha}\psi_b(ht^{-y_0\nu}) \\ \nonumber
f_s & \sim & t^{2-\alpha_s}\psi_s(ht^{-y_0\nu},h_1t^{-y_1\nu}) \\ \nonumber
f_e & \sim & t^{2-\alpha_e}\psi_e(ht^{-y_0\nu},h_1t^{-y_1\nu},h_2t^{-y_2\nu}), 
\end{eqnarray}
where $t$ is the reduced temperature $(T-T_c)/T_c;$ $y_0,$ $y_1$ and $y_2$ are the bulk, surface and edge scaling indices, from which all the susceptibility critical exponents follow. The reduced magnetic fields in the bulk, surface and edge are denoted, respectively $h,$ $h_1$ and $h_2.$ In particular, the bulk susceptibility is given by $$\chi = \partial^2f_b/\partial h^2 \asymp t^{-\gamma}; \,\,\, \gamma=\nu(2y_0-d),$$ the surface susceptibilities are given by
$$\chi_{1} = \partial^2f_s/\partial h \partial h_1 \asymp t^{-\gamma_{1}},\qquad \gamma_{1}=\nu(y_0+y_1-d+1),$$
$$\chi_{11} = \partial^2f_s/\partial h_1^2  \asymp t^{-\gamma_{11}},\qquad \gamma_{11}=\nu(2y_1-d+1),$$
and the edge susceptibilities are given by
$$\chi_{2} = \partial^2f_e/\partial h \partial h_2 \asymp t^{-\gamma_{2}},\qquad \gamma_{2}=\nu(y_0+y_2-d+2),$$
$$\chi_{21} = \partial^2f_e/\partial h_1 \partial h_2  \asymp t^{-\gamma_{21}},\qquad \gamma_{21}=\nu(y_1+y_2-d+2).$$
In \cite{GT84} it was shown (non-rigorously) that, for the O$(n=0)$ model, $y_2(\alpha) = -5\pi/8\alpha,$ where $\alpha$ is the wedge angle. Similarly, in \cite{BPP84} it was shown (non-rigorously) that, for the O$(n=1)$ model, $y_2(\alpha) = -\pi/2\alpha.$ For the Gaussian model, corresponding to the O$(n=-2)$ model, $y_2(\alpha) = -\pi/\alpha.$  Thus for these three cases, we have $$y_2(\alpha) = -\pi\sigma/\alpha.$$ 
If, as we conjecture, this is true for other values of $n,$ we then find
$$\gamma_2 = \frac{\kappa^2+8\kappa+12-24\pi/\alpha+4\pi \kappa/\alpha}{8\kappa(4-\kappa)},$$
and
$$\gamma_{21} = \frac{3\kappa-6-6\pi/\alpha+\pi \kappa/\alpha}{2\kappa(4-\kappa)}.$$
where we have used $\sigma=\frac{3}{\kappa}-\frac{1}{2}$. If we set $\alpha = \pi,$ they reduce to (\ref{gam1}), providing evidence for the validity of our assumption that they hold for all $n \in [-2,2).$ For $n=2$ the free energy is believed to exhibit an essential singularity, so that critical exponents do not exist. This is signalled in the conjectured exponent values by the divergence at $\kappa=4,$ corresponding to the O$(n=2)$ model.

We remark that the scaling indices $y_0,$ $y_1$ and $y_2$ take  particularly simple forms if parameterised in terms of $\sigma$: 
$$y_0=\frac{(\sigma+1)(\sigma+2)}{2\sigma+1},\qquad y_1=1-\sigma,\qquad y_2=-\frac{\pi \sigma}{\alpha}.$$

%%%%%%%%%%%%%%%%%%%%%%%%%%%%%%
\section{Conclusion}
%%%%%%%%%%%%%%%%%%%%%%%%%%%%%%
%It seems to us remarkable that the critical exponent that enters into the winding distribution should be expressible in terms of half-plane bulk and surface critical exponents. The agreement between the predicted value of this exponent and the individually predicted values of the surface  exponents provides additional reassurance as to the correctness of all predicted exponent values.
We have generalised the identity of Duminil-Copin and Smirnov off criticality, which allows us to make statements about critical exponents. We have proved an inequality for surface and wedge exponents, subject only to their existence.

We have similarly proved, under the same assumption, a relationship between the surface susceptibility exponents and the winding angle exponent of the O($n$) model for $n\in[-2,2)$. Previously conjectured values of the surface and winding angle exponents are in agreement with the relationship we have derived for all values of $n\in[-2,2)$. 

A study of the edge exponents that arise when considering the O$(n)$ model in a wedge geometry permits us to conjecture the exact value of the exponents for all wedge angles. 

The off-critical extension of Smirnov's identity that we have obtained seems likely to yield a number of other significant results, and we are currently studying this possibility.

\section*{Acknowledgment}
We are grateful for financial support from the Australian Research Council (ARC). This work was carried out  during the visit of three of the authors to the US Mathematical Sciences Research Institute (MSRI, USA), during the Spring 2012 Random Spatial Processes Program. The authors thank the institute for its hospitality and the NSF (grant DMS-0932078) for its financial support. AJG wishes to thank Neal Madras and Mireille Bousquet-M$\mathrm{\acute{e}}$lou for helpful discussions.

\end{document}